\documentclass[twocolumn,aps,superscriptaddress,showpacs,floatfix,longbibliography]{revtex4-1}
\usepackage{url}
\usepackage{cancel}
\usepackage[colorlinks,linkcolor=blue,citecolor=blue,filecolor=black,urlcolor=blue]{hyperref}
\usepackage{epsfig,graphics}
\usepackage{graphicx}
\usepackage{dcolumn}
\usepackage{bm}
\usepackage[usenames]{color}
\usepackage{ulem} 
\usepackage{amssymb}
\usepackage{amsmath}
\usepackage{multirow}
\usepackage{float}
\usepackage{MnSymbol}
\usepackage{appendix}
\usepackage{color}
\usepackage{cleveref}
\usepackage{mathtools}
\usepackage{dsfont} 

\begin{document}
\title{Three-dimensional QCD phase diagram in the pNJL model}
\author{Lu-Meng Liu}
\affiliation{School of Physical Sciences, University of Chinese Academy of Sciences, Beijing 100049, China}
\author{Jun Xu}\email[Correspond to ]{junxu@tongji.edu.cn}
\affiliation{School of Physics Science and Engineering, Tongji University, Shanghai 200092, China}
\affiliation{Shanghai Advanced Research Institute, Chinese Academy of Sciences, Shanghai 201210, China}
\affiliation{Shanghai Institute of Applied Physics, Chinese Academy of Sciences, Shanghai 201800, China}
\author{Guang-Xiong Peng}
\affiliation{School of Nuclear Science and Technology, University of Chinese Academy of Sciences, Beijing 100049, China}
\date{\today}

\begin{abstract}
Based on the three-flavor Polyakov-looped Nambu$-$Jona-Lasinio (pNJL) model, we have studied the structure of the three-dimensional QCD phase diagram with respect to the temperature, the baryon chemical potential, and the isospin chemical potential, by investigating the interplay among the chiral quark condensate, the pion condensate, and the Polyakov loop. While the pNJL model leads to qualitatively similar structure of the normal quark phase, the pion superfluid phase, and the Sarma phase as well as their phase boundaries, when compared to the NJL model, the inclusion of the Polyakov loop enlarges considerably the areas of the pion superfluid phase and the Sarma phase, and leads to critical end points at higher temperatures. With the contribution of the gluon dynamics effectively included, the present study is expected to give a more reliable prediction of the three-dimensional QCD phase diagram compared to that in the NJL model.
\end{abstract}
\maketitle


\section{Introduction}
\label{sec:intro}

Exploring the phase diagram of the quantum chromodynamics (QCD) has been a main task in high-energy nuclear physics over the past few decades. Although the lattice QCD (LQCD) calculations favor a smooth crossover from the hadronic phase to the partonic phase at high temperatures and small baryon chemical potentials~\cite{Bernard:2004je,Aoki:2006we, Bazavov:2011nk}, they suffer from the sign problem~\cite{Karsch:2001cy, Muroya:2003qs, Bedaque:2017epw} at large baryon chemical potentials, where our knowledge on the QCD phase diagram mostly relies on experimental data from heavy-ion collisions at RHIC-BES, FAIR-CBM, NICA, and HIAF, etc., as well as theoretical studies based on effective QCD models. The latter includes the NJL model~\cite{Bratovic:2012qs,Asakawa:1989bq, Fukushima:2008wg, Carignano:2019ivp}, the Dyson-Schwinger (DS) equation approach~\cite{Xin:2014ela, Fischer:2014ata}, the functional renormalization group (FRG) method~\cite{Fu:2019hdw,Gao:2020qsj}, and the quark-meson coupling model~\cite{Frasca:2011zn, Herbst:2010rf, Schaefer:2009ui}, etc. Besides the baryon chemical potential and the temperature, our knowledge on the QCD phase diagram can be extended to other degree of freedom, e.g., the isospin~\cite{Son:2000xc}. If the isospin chemical potential exceeds the mass of a pion, pions can be produced out of the vacuum, and the resulting pion condensate may dominate the QCD phase structure at large isospin chemical potentials~\cite{Liu:2021gsi, Klein:2003fy, Barducci:2004tt, Barducci:2005ut, He:2005nk, Ebert:2005cs, Xia:2013caa, Roessner:2006xn, Zhang:2006dn, Zhang:2006gu, Sasaki:2010jz, Mu:2010zz, Adhikari:2018cea, Lu:2019diy, Brandt:2017oyy,Khunjua:2019nnv}.

In the previous study, we have obtained the QCD phase diagram at finite temperatures, baryon chemical potentials, and isospin chemical potentials in the three-flavor NJL model~\cite{Liu:2021gsi} with the scalar-isovector and vector-isovector couplings. Typically, we have fitted the coupling constants of the scalar-isovector and vector-isovector interactions by reproducing the physical pion mass and the isospin density from LQCD calculations in baryon-free quark matter, and then extrapolated the calculations to finite baryon chemical potentials. While the NJL model has the advantage of describing chiral phase transitions, it is well-known that this model lack of gluon dynamics and is unable to describe the deconfinement phase transition. For this reason, it gives a lower temperature of the QCD critical end point (CEP), compared to that obtained from the DS equation approach and the FRG method. To overcome this drawback, one needs to introduce the Polyakov loop into the NJL model~\cite{Fukushima:2003fw, Fukushima:2008wg, Ratti:2005jh}, leading to the so-called pNJL model. The Polyakov loop $\Phi~(\bar{\Phi})$ is related to the excess free energy for a static quark (anti-quark) in a hot gluon medium~\cite{Fukushima:2010bq}, and thus serving as an order parameter for the deconfinement phase transition which is characterized by the spontaneous breaking of the $Z(N_c)$ center symmetry of $\mathrm{QCD}$. Exploring the three-dimensional QCD phase diagram based on the pNJL model is helpful for understanding the interplay among different order parameters, e.g., the chiral condensate, the pion condensate, and the Polyakov loop, and mapping out the resulting detailed phase structures.


\section{Theoretical framework}
\label{sec:theory}
We start from the following Lagrangian density of the three-flavor pNJL model~\cite{Fukushima:2017csk, Costa:2008dp, Fu:2007xc}
\begin{eqnarray} \label{PNJL Lagrangian}
	\mathcal{L}_\mathrm{pNJL}
	&=& \bar{\psi}(i\gamma^\mu D_\mu+\hat{\mu}\gamma^0-\hat{m})\psi+ \mathcal{L}_\mathrm{S} + \mathcal{L}_\mathrm{V} \nonumber\\
	&& + \mathcal{L}_{\mathrm{KMT}} + \mathcal{L}_\mathrm{IS} + \mathcal{L}_\mathrm{IV}-\mathcal{U}(\Phi,\bar{\Phi},T),
\end{eqnarray}
where
\begin{eqnarray}
	\mathcal{L}_\mathrm{S}&=& \frac{G_\mathrm{S}}{2}\sum_{a=0}^8 [(\bar{\psi}\lambda^a\psi)^2 + (\bar{\psi}i \gamma^5 \lambda^a \psi)^2], \\
	\mathcal{L}_\mathrm{V}&=& - \frac{G_\mathrm{V}}{2}\sum_{a=0}^8 [(\bar{\psi}\gamma^{\mu}\lambda^a\psi)^2 + (\bar{\psi} \gamma^5 \gamma^{\mu} \lambda^a \psi)^2],\\
	\mathcal{L}_{\mathrm{KMT}}&=& -K[\mathrm{det}\bar{\psi}(1+\gamma^5)\psi + \mathrm{det}\bar{\psi}(1-\gamma^5)\psi],
\end{eqnarray}
\begin{eqnarray}
	\mathcal{L}_\mathrm{IS}&=& G_\mathrm{IS} \sum_{a=1}^3[(\bar{\psi}\lambda^a\psi)^2 + (\bar{\psi}i \gamma^5 \lambda^a \psi)^2], \\
	\mathcal{L}_\mathrm{IV}&=& - G_\mathrm{IV} \sum_{a=1}^3[(\bar{\psi}\gamma^{\mu}\lambda^a\psi)^2 + (\bar{\psi} \gamma^5 \gamma^{\mu} \lambda^a \psi)^2],
\end{eqnarray}
are the scalar-isoscalar term, the vector-isoscalar term, the Kobayashi-Maskawa-t'Hooft (KMT) term, the scalar-isovector term, and the vector-isovector term, respectively. In the above, $\psi = (u,d,s)^T$ represents the three-flavor quark fields with each flavor containing quark fields of three colors; $\hat{\mu} = \text{diag}(\mu_u, \mu_d, \mu_s)$ and $\hat{m}=\text{diag}(m_u,m_d,m_s)$ are the matrices of the chemical potential and the current quark mass for $u$, $d$, and $s$ quarks;  $D_\mu=\partial_\mu - i A_\mu$ is the covariant derivative with $A_\mu= \delta_{\mu}^0 A_0$, where $A_0= \textsl{g} A_0^a \lambda^a/2 =-iA_4$ is the non-Abelian $\mathrm{SU}(3)$ gauge field with the gauge coupling $\textsl{g}$ conveniently absorbed in the definition of $A_\mu$; $\lambda^a $ ($a=1,...,8$) are the Gell-Mann matrices in $\mathrm{SU}(3)$ flavor space with $\lambda^0 =\sqrt{2/3}\mathds{1}_3$; $G_\mathrm{S}$ and $G_\mathrm{V}$ are respectively the scalar-isoscalar and the vector-isoscalar coupling constant; $G_\mathrm{IS}$ and $G_\mathrm{IV}$ are respectively the scalar-isovector and the vector-isovector coupling constant. Since the Gell-Mann matrices with $a=1, 2, 3$ are identical to the Pauli matrices in $u$ and $d$ space, the isovector couplings break the $\mathrm{SU}(3)$ symmetry while keeping the isospin symmetry. $K$ denotes the strength of the six-point KMT interaction \cite{tHooft:1976snw} that breaks the axial $U(1)_\mathrm{A}$ symmetry, where `det' denotes the determinant in flavor space. In the present study, we employ the parameters $m_u = m_d = 3.6$ MeV, $m_s = 87$ MeV, $G_\mathrm{S}\Lambda^2 = 3.6$, $K\Lambda^5 = 8.9$, and the cutoff value in the momentum integral $\Lambda = 750$ MeV/c given in Refs.~\cite{Lutz:1992dv,Buballa:2003qv,Bratovic:2012qs}. In our previous study~\cite{Liu:2021gsi}, $G_\mathrm{IS} =-0.002 G_\mathrm{S}$ and $G_\mathrm{IV} = 0.25G_\mathrm{S}$ are determined by fitting the physical pion mass $m_\pi \approx 140.9$ MeV and the reduced isospin density from LQCD calculations at zero temperature~\cite{Brandt:2018bwq}, at which the pNJL model reduces to the NJL model. We set $G_\mathrm{V}=0$ and $\mu_s=0$ throughout the present study.

We take the temperature-dependent effective potential $\mathcal{U}(\Phi,\bar{\Phi},T)$ from Ref.~\cite{Fukushima:2008wg}, i.e.,
\begin{eqnarray}
\mathcal{U}(\Phi,\bar{\Phi},T) &=&
    -b \cdot T\{54e^{-a/T}\Phi\bar{\Phi} +\ln[1-6\Phi\bar{\Phi}  \nonumber\\
&&  -3(\Phi\bar{\Phi})^2+4(\Phi^3+\bar{\Phi}^3)]\}.
\end{eqnarray}
The parameters $a=664$ MeV and $b=0.028\Lambda^3$ are determined by the condition that the first-order phase transition in the pure gluodynamics takes place at $T = 270$ MeV~\cite{Fukushima:2008wg}, and the simultaneous crossover of the chiral restoration and the deconfinement phase transition occurs around $T\approx 212$ MeV. The Polyakov loop $\Phi$ and its (charge) conjugate $\bar{\Phi}$ are expressed as~\cite{Pisarski:2000eq,Ratti:2005jh}
\begin{equation}\label{eq:Philoop}
	\Phi = \frac{1}{N_c}\mathrm{Tr}_c L,~~~~~~\bar{\Phi} = \frac{1}{N_c}\mathrm{Tr}_c L^{\dagger},
\end{equation}
where $N_c=3$ is the color degeneracy, and the matrix $L$ in color space is explicitly given by
\begin{equation}
	L(\vec{x}) = \mathcal{P}\mathrm{exp}\left[i\int_0^{\beta}d\tau A_4(\tau, \vec{x}) \right] = \mathrm{exp}\left(\frac{iA_4}{T}\right),
\end{equation}
with $\mathcal{P}$ being the path ordering and $\beta=1/T$ being the inverse of temperature. The coupling between the Polyakov loop and quarks is uniquely determined by the covariant derivative $D_\mu$ in the pNJL Lagrangian [Eq.~(\ref{PNJL Lagrangian})]~\cite{Ratti:2005jh}. The second equal sign in the above equation is valid by treating the temporal component of the Euclidean gauge field $A_4$ as a constant in the pNJL model. In this way, the Polyakov loop $\Phi$ and its conjugate $\bar{\Phi}$ can be treated as classical field variables.

Based on the mean-field approximation, the Lagrangian density of the pNJL model can be written as
\begin{equation}\label{eq:MFLagrangian}
	\mathcal{L}_\mathrm{MF}=\bar{\psi}\mathcal{S}^{-1}\psi -\mathcal{V}-\mathcal{U}(\Phi,\bar{\Phi},T),
\end{equation}
where
\begin{widetext}
\begin{eqnarray}\label{eq:propagator}
	&&\mathcal{S}^{-1} (p) =
	\begin{pmatrix}
		\gamma^\mu p_\mu + \left(\frac{\tilde{\mu}_\mathrm{B}}{3}- iA_4 + \frac{\tilde{\mu}_\mathrm{I}}{2}\right) \gamma^0-M_u	 &	i \Delta \gamma^5 	& 0
		\\
		i \Delta\gamma^5 	&\gamma^\mu p_\mu +\left(\frac{\tilde{\mu}_\mathrm{B}}{3}- iA_4 - \frac{\tilde{\mu}_\mathrm{I}}{2}\right) \gamma^0 -M_d	&	0
		\\
		0& 0 &	\gamma^\mu p_\mu + \left(\frac{\tilde{\mu}_\mathrm{B}}{3}- iA_4 - {\tilde{\mu}_\mathrm{S}}\right)\gamma^0 -M_s
	\end{pmatrix}
\end{eqnarray}
\end{widetext}
is the inverse of the quark propagator $\mathcal{S}(p)$ as a function of quark momentum $p$, with
\begin{equation}\label{Delta}
\Delta = \left(G_\mathrm{S} + 2G_\mathrm{IS} -K \sigma_s\right)\pi
\end{equation}
being the gap parameter, and
\begin{eqnarray}
	\mathcal{V} &=&  G_\mathrm{S} \left( \sigma_u^2 + \sigma_d^2  + \sigma_s^2 \right)  + \frac{G_\mathrm{S}}{2}\pi^2 + G_\mathrm{IS} (\sigma_u-\sigma_d)^2  \nonumber\\
	&+& G_\mathrm{IS}\pi^2 - 4K \sigma_u \sigma_d \sigma_s - K\sigma_s \pi^2 \nonumber\\
	&-&\frac{1}{3}G_\mathrm{V} \left( \rho_u+\rho_d+\rho_s \right)^2 -G_\mathrm{IV}(\rho_{u}-\rho_{d})^2
\end{eqnarray}
being the condensation energy independent of the quark fields. In the above, $\rho_q=\langle \bar{q} \gamma^0 q \rangle$ and $\sigma_q =\langle \bar{q} q \rangle$ are the net-quark density and the chiral condensate, respectively, with $q=u,d,s$ being the quark flavor, and $\pi=\langle \bar{\psi} i \gamma^5 \lambda^1 \psi \rangle$ is the pion condensate. The constituent mass of quarks can be expressed as
\begin{eqnarray}
	M_u &=& m_u-2G_\mathrm{S}\sigma_u - 2G_\mathrm{IS}(\sigma_u-\sigma_d) + 2K\sigma_d \sigma_s ,\nonumber\\
	M_d &=& m_d-2G_\mathrm{S}\sigma_d + 2G_\mathrm{IS}(\sigma_u-\sigma_d) + 2K\sigma_u \sigma_s ,\nonumber\\
	M_s &=& m_s-2G_\mathrm{S}\sigma_s + 2K\sigma_u \sigma_d + \frac{K}{2} \pi^2.\nonumber
\end{eqnarray}
The effective baryon, isospin, and strangeness chemical potentials are defined as
\begin{eqnarray}\label{eff}
	\tilde{\mu}_\mathrm{B} &=& \mu_\mathrm{B} - 2G_\mathrm{V}\rho, \nonumber\\
	\tilde{\mu}_\mathrm{I} &=& \mu_\mathrm{I} - 4G_\mathrm{IV}\left(\rho_u-\rho_d\right) , \nonumber\\
	\tilde{\mu}_\mathrm{S} &=& \mu_\mathrm{S},
\end{eqnarray}
with
\begin{eqnarray}
	\mu_\mathrm{B} &=& \frac{3(\mu_u + \mu_d)}{2} , \nonumber\\
	\mu_\mathrm{I} &=& \mu_u - \mu_d, \nonumber\\
	\mu_\mathrm{S} &=& \frac{\mu_u + \mu_d}{2} - \mu_s.
\end{eqnarray}

The thermodynamic potential of the quark system can be obtained through
\begin{eqnarray}\label{eq:Omega1}
	\Omega = -T\sum_n\int \frac{d^3 p}{(2\pi)^3} \text{Tr ln } \mathcal{S}(i\omega_n, \vec{p})^{-1} +\mathcal{V}+\mathcal{U}(\Phi,\bar{\Phi},T). \nonumber\\
\end{eqnarray}
In the above, the four-momentum $p = (p_0, \vec{p})$ becomes $p=(i\omega_n, \vec{p})$ with $\omega_n = (2n+1)\pi T$ being the Matsubara frequency for a Fermi system. In order to evaluate $\Omega$ for each momentum $p$ numerically, we need to find the zeros of $\mathcal{S}^{-1} (p)$. Similar to the method in Refs.~\cite{Steiner:2002gx,Fukushima:2004zq,Ruester:2005jc}, it can be proved that the eigenvalues $\lambda_k$ $(k=1,2,3,4)$ of the following ``Dirac Hamiltonian density"
\begin{eqnarray}
	\mathcal{H} (\vec{p})
	& =&-\begin{pmatrix}
		\frac{\tilde{\mu}_\mathrm{I}}{2} -M_u	 &	|\vec{p}| 	& 0 & -\Delta
		\\
		|\vec{p}|& \frac{\tilde{\mu}_\mathrm{I}}{2} +M_u	& \Delta &	0
		\\
		0& 	\Delta &	-\frac{\tilde{\mu}_\mathrm{I}}{2} -M_d & |\vec{p}|
		\\
		-\Delta&  0	 &	|\vec{p}| & -\frac{\tilde{\mu}_\mathrm{I}}{2} +M_d
	\end{pmatrix} \nonumber
\end{eqnarray}
are zeros of $\mathcal{S}^{-1} (p)$. Using the relationship $\mathrm{Trln}=\mathrm{lnDet}$, one can get the following expression of the thermodynamic potential
\begin{eqnarray}\label{eq:Omega}
	\Omega 	&=& \Omega^+(\lambda'_1)+\Omega^+(\lambda'_2)+\Omega^-(-\lambda'_3)+\Omega^-(-\lambda'_4) \nonumber\\
	&&+\Omega^+(E_s^-)+\Omega^-(E_s^+)+\mathcal{V}+\mathcal{U}(\Phi,\bar{\Phi},T)
\end{eqnarray}
with
\begin{eqnarray} \label{eq:Omega-pm}
	\Omega^{\pm}(\lambda) &=& -2N_c\,\int_0^{\Lambda}  \frac{d^3 p}{(2\pi)^3}\frac{\lambda}{2} -2T\int_0^{\Lambda} \frac{d^3 p}{(2\pi)^3}Z^{\pm}(-\lambda), \nonumber\\
\end{eqnarray}
where the integrands in the second integral are
\begin{eqnarray}
    Z^{-}(\lambda) &=&	\mathrm{Tr}_c\text{ln}\left( 1+L \xi_{\lambda}\right) = \text{ln} \left\{ 1+N_c\Phi \xi_{\lambda}+N_c\bar{\Phi}\xi_{\lambda}^2 + \xi_{\lambda}^3 \right\}, \nonumber\\
	Z^{+}(\lambda) &=&\mathrm{Tr}_c\text{ln}\left( 1+L^\dagger \xi_{\lambda}\right) =\text{ln} \left\{  1+N_c\bar{\Phi} \xi_{\lambda}+N_c{\Phi}\xi_{\lambda}^2 + \xi_{\lambda}^3 \right\}, \nonumber
\end{eqnarray}
with $\xi_{\lambda}=e^{\beta \lambda}$. In Eq.~(\ref{eq:Omega}), $\lambda'_k$ are defined as $\lambda'_k = \lambda_k-\frac{\tilde{\mu}_\mathrm{B}}{3}$, and $E_s^{\pm}= E_s \pm \tilde{\mu}_s$ with $E_s=\sqrt{M_s^2 + \vec{p}^2}$ is the single $s$ quark energy. Throughout this paper, $\mathrm{Tr}$ and $\mathrm{Det}$ represent respectively the trace and determinant over Dirac, flavor, and color space, while $\mathrm{Tr}_c$ and $\mathrm{Det}_c$ represent those only taken over color space. It should be pointed out that we introduce a momentum cutoff in the two integrals in Eq.~(\ref{eq:Omega-pm}) as in Ref.~\cite{Costa:2008dp}, otherwise the integrals will be divergent at large baryon and isospin chemical potentials. This is, however, slightly different from our previous studies~\cite{Liu:2016yid,Liu:2021bzf,PhysRevC.104.044901}.

By taking the trace of the corresponding component of the propagator~\cite{He:2005nk}, the chiral condensates, the net-quark densities, and the pion condensate can be expressed as
\begin{eqnarray}\label{eq:gap equation}
	\sigma_u
	&=&  4\,  N_c\sum_{k=1}^4\int \frac{d^3 p}{(2\pi)^3}g_{\sigma u}\left(\lambda_k\right) \left[-\frac{1}{2}+F^+\left(\lambda'_k\right)\right],\\
	\sigma_d
	&=& 4\,  N_c\sum_{k=1}^4\int \frac{d^3 p}{(2\pi)^3}g_{\sigma d}\left(\lambda_k\right) \left[-\frac{1}{2}+F^+\left(\lambda'_k\right)\right],\\
	\sigma_{s}
	&=& 2\,N_c\int \frac{d^3 p}{(2\pi)^3}\frac{M_s}{E_s} \left[F^+\left(E_s^-\right)+F^-\left(E_s^+\right)-1\right], \\
	\rho_u
	&=& 4\,  N_c\sum_{k=1}^4\int \frac{d^3 p}{(2\pi)^3}g_{\rho u}\left(\lambda_k\right) \left[-\frac{1}{2}+F^+\left(\lambda'_k\right)\right],\\
	\rho_d
	&=& 4\,  N_c\sum_{k=1}^4\int \frac{d^3 p}{(2\pi)^3}g_{\rho d}\left(\lambda_k\right) \left[-\frac{1}{2}+F^+\left(\lambda'_k\right)\right], \\
	\rho_{s}
	&=&2\,N_c\int \frac{d^3 p}{(2\pi)^3}\left[F^+\left(E_s^-\right)-F^-\left(E_s^+\right)\right],\\
	\pi
	&=& 4\,N_c \sum_{k=1}^4\int \frac{d^3 p}{(2\pi)^3}g_{\pi}\left(\lambda_k\right) \left[-\frac{1}{2}+F^+\left(\lambda'_k\right)\right],\label{eq:pion}
\end{eqnarray}
where the $g$ functions have the same form as those in Ref.~\cite{Liu:2021gsi}, since the $g$ functions are actually independent of $\tilde{\mu}_\mathrm{B}$ and the $iA_4$ terms are always combined with ${\tilde{\mu}_\mathrm{B}}/{3}$ in Eq.~(\ref{eq:propagator}). In the above, $F^+(\lambda)$ and $F^-(\lambda)$ are, respectively, the effective phase-space distribution for quarks and antiquarks, and they are expressed as
\begin{eqnarray}
	F^+(\lambda)
	&=& \frac{1}{N_c}\mathrm{Tr}_c\left( \frac{1}{1+L \xi_{\lambda}}\right)
	=\frac{1+2\Phi \xi_{\lambda}+ \bar{\Phi}\xi_{\lambda}^2}
	{ 1+N_c\Phi \xi_{\lambda}+N_c\bar{\Phi}\xi_{\lambda}^2 + \xi_{\lambda}^3}, \nonumber\\
	F^-(\lambda)
	&=& \frac{1}{N_c}\mathrm{Tr}_c\left( \frac{1}{1+L^\dagger \xi_{\lambda}}\right)
	=\frac{1+2\bar{\Phi} \xi_{\lambda}+ \Phi \xi_{\lambda}^2}
	{ 1+N_c\bar{\Phi} \xi_{\lambda}+N_c\Phi \xi_{\lambda}^2 + \xi_{\lambda}^3}. \nonumber
\end{eqnarray}
It is seen that the above distributions reduce to the normal Fermi-Dirac form at high temperatures when the Polyakov loops are approaching 1, while they become the Fermi-Dirac form with a reduced temperature of $T/3$ at low temperatures when the Polyakov loops are almost zero. This leads to a CEP at a higher temperature in the pNJL model than in the NJL model. Equations~(\ref{eq:gap equation})-(\ref{eq:pion}) can be also obtained equivalently from
\begin{eqnarray}
\frac{\partial \Omega}{\partial \sigma_q} = \frac{\partial \Omega}{\partial \rho_q} = \frac{\partial \Omega}{\partial \pi} = 0,
\end{eqnarray}
with $q=u,d,s$ being the quark flavor, leading to the relations
\begin{eqnarray}
\sigma_q =\frac{\partial \Omega}{\partial M_q},~\rho_q = -\frac{\partial \Omega}{\partial \mu_q},~\pi=-\frac{\partial \Omega}{\partial \Delta}.
\end{eqnarray}
The values of $\Phi$ and $\bar{\Phi}$ can be similarly determined by minimizing the grand potential with respect to the Polyakov loops, i.e.,
\begin{eqnarray}\label{pomega}
	\frac{\partial \Omega}{\partial \Phi}  = \frac{\partial \Omega}{\partial \bar{\Phi}}  =0.
\end{eqnarray}

\section{Results and discussions}
\label{sec:results}

With the theoretical framework described above, we will display how the order parameters, i.e., the chiral condensate, the pion condensate, and the Polyakov loop, evolve with the baryon chemical potential, the isospin chemical potential, and the temperature. After discussing the interplay among these order parameters, we will then present the three-dimensional QCD phase diagram. Results in the present study based on the pNJL model will also be compared with those from the NJL model. In the NJL model with the same parameter values, the Polyakov-loop potential is turned off, and the covariant derivative $D_\mu$ is reduced to $\partial_\mu$ in Eq.~(\ref{PNJL Lagrangian}).

\subsection{Interplay among order parameters}

\begin{figure}[ht]
	\includegraphics[scale=0.35]{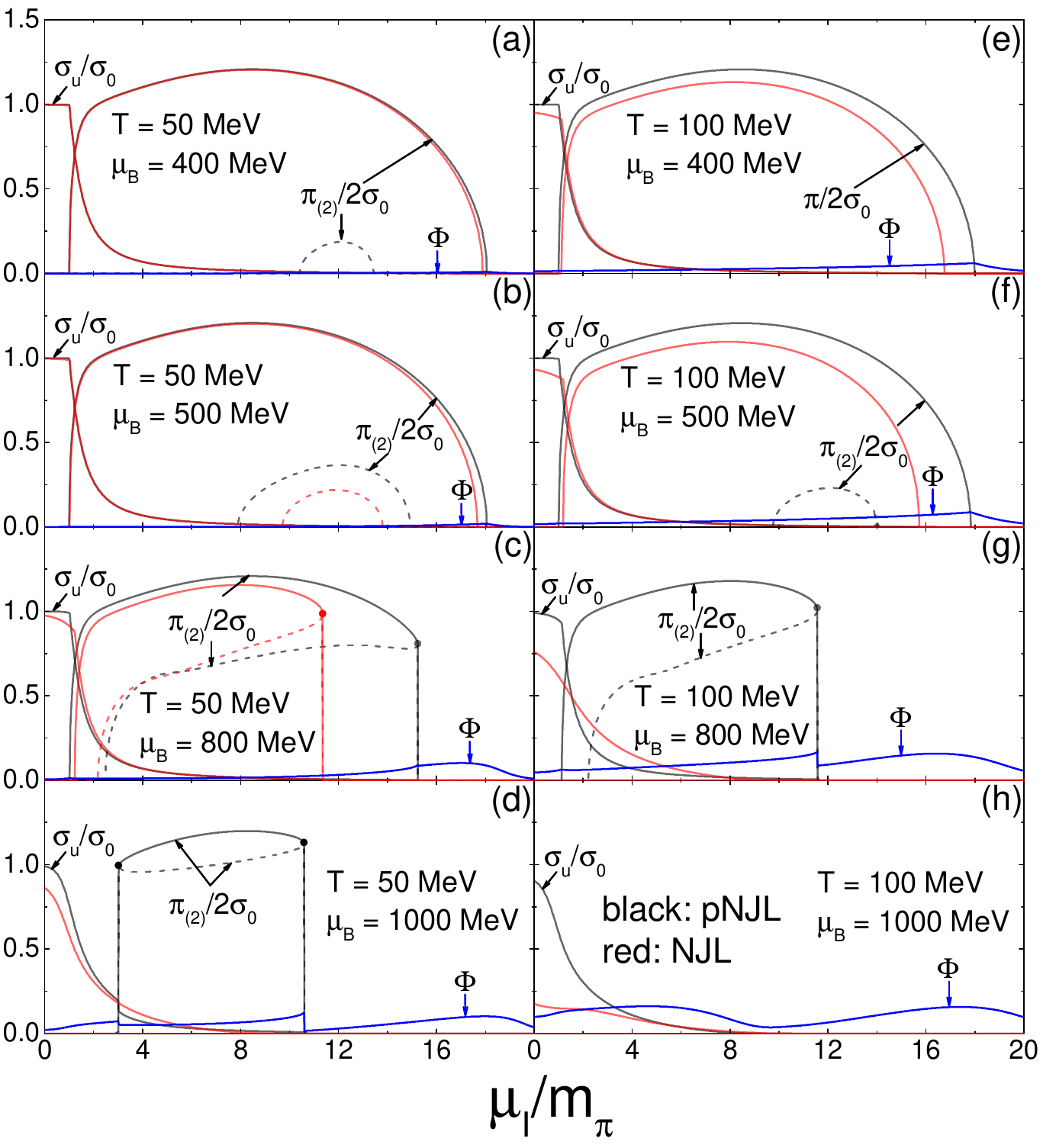}
	\caption{(Color online) Reduced pion condensate $\pi/2\sigma_0$ and chiral condensate $\sigma_u/\sigma_0$ as well as the Polyakov loop $\Phi$ as a function of the reduced isospin chemical potential $\mu_\mathrm{I}/m_\pi$ in hot [$T=50$ (left) and $100$ (right) MeV] and baryon-rich [$\mu_\mathrm{B}=400$ (a,e), 500 (b,f), 800 (c,g), and 1000 (d,h) MeV] quark matter. Results are compared with those obtained from the NJL model.}\label{fig:pion-T=100}
\end{figure}

We compare the pion and chiral condensates in baryon-rich quark matter at $T=50$ and 100 MeV as a function of the isospin chemical potential based on the NJL and pNJL model in Fig.~\ref{fig:pion-T=100}. One sees that the pion condensate appears around $\mu_\mathrm{I} \sim m_\pi$ and disappears at very large $\mu_\mathrm{I}$ or high temperatures. The appearance and disappearance of the pion condensate are second-order phase transitions at smaller $\mu_\mathrm{B}$. With the increasing $\mu_\mathrm{B}$, the disappear of the pion condensate first becomes a first-order phase transition, and then the appearance of the pion condensate becomes a first-order one as well. At very large $\mu_\mathrm{B}$, there is no pion condensate. At intermediate $\mu_\mathrm{B}$, there exists a second nonzero solution $\pi_2$ (denoted as dashed lines), which corresponds to the local maximum of the thermodynamic potential and is called the Sarma phase~\cite{Sarma:1963a} as detailed in Ref.~\cite{Liu:2021gsi}. It is seen that the pion superfluid phase ($\pi$) and the Sarma phase ($\pi_2$) exist in a broader region of chemical potentials and temperatures in the pNJL model than in the NJL model.

\begin{figure}[ht]
	\includegraphics[scale=0.3]{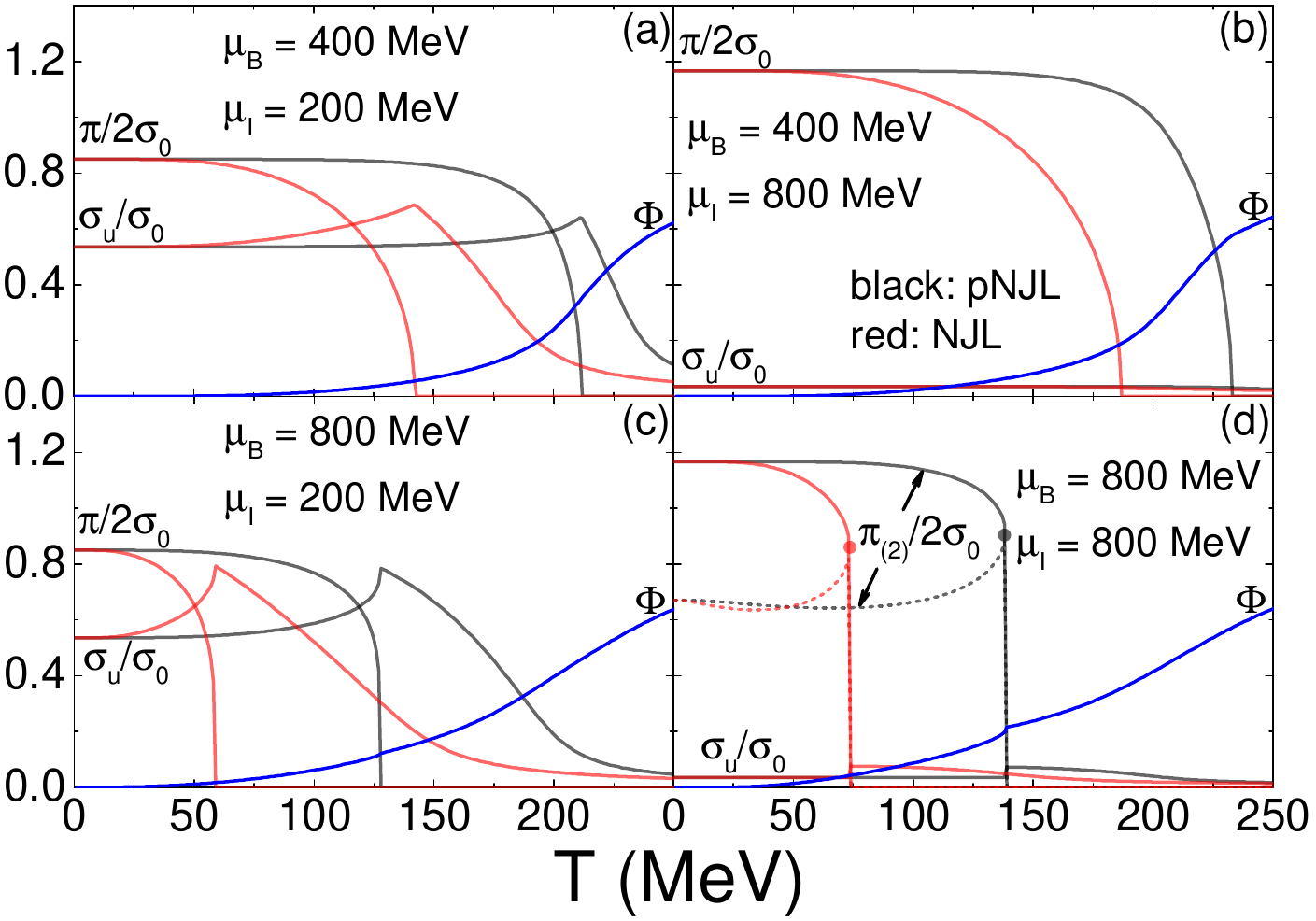}
	\caption{(Color online)  Reduced pion condensate $\pi/2\sigma_0$ and chiral condensate $\sigma_u/\sigma_0$ as well as the Polyakov loop $\Phi$ as a function of the temperature $T$ in quark matter of different baryon chemical potentials $\mu_\mathrm{B}$ and isospin chemical potentials $\mu_\mathrm{I}$. Results are compared with those obtained from the NJL model.}\label{fig:pion-T}
\end{figure}

Figure~\ref{fig:pion-T} displays the temperature dependence of the pion and chiral condensates at various baryon and isospin chemical potentials. One sees that the difference in the behavior of the pion condensate between the pNJL model and the NJL model is mainly at high temperatures. The temperatures of the CEP for the disappearance of pion condensates $\pi_{(2)}$ in the pNJL model are much higher than that in the NJL model, which will manifest themselves in the results of the phase diagrams to be shown later. Again, the pion superfluid phase ($\pi$) and the Sarma phase ($\pi_2$) exist in a broader region of temperatures in the pNJL model than in the NJL model.

Some common features in Figs.~\ref{fig:pion-T=100} and \ref{fig:pion-T} need further discussions. According to the expressions of $g$ functions in Ref.~\cite{Liu:2021gsi} and the gap equations [Eqs.~(\ref{eq:gap equation})-(\ref{eq:pion})], both the chiral condensate and the net-quark densities depend on the gap parameter $\Delta$ or equivalently the pion condensate $\pi_{(2)}$. Therefore, the sudden change of $\pi_{(2)}$, corresponding to either a first-order or a second-order phase transition of the whole quark matter system, leads to a sudden jump of the chiral condensates, the net-number densities as well as the Polyakov loop $\Phi$. For the behavior of the Polyakov loop $\Phi$, in principle one expects that it should increase with both the increasing chemical potential and temperature, while the non-monotonical dependence of $\Phi$ on $\mu_\mathrm{I}$ in Fig.~\ref{fig:pion-T=100} is due to the momentum cutoff in Eq.~(\ref{eq:Omega-pm}) when evaluating Eq.~(\ref{pomega}).

\subsection{Three-dimensional QCD phase diagram}

\begin{figure}[ht]
	\includegraphics[scale=0.3]{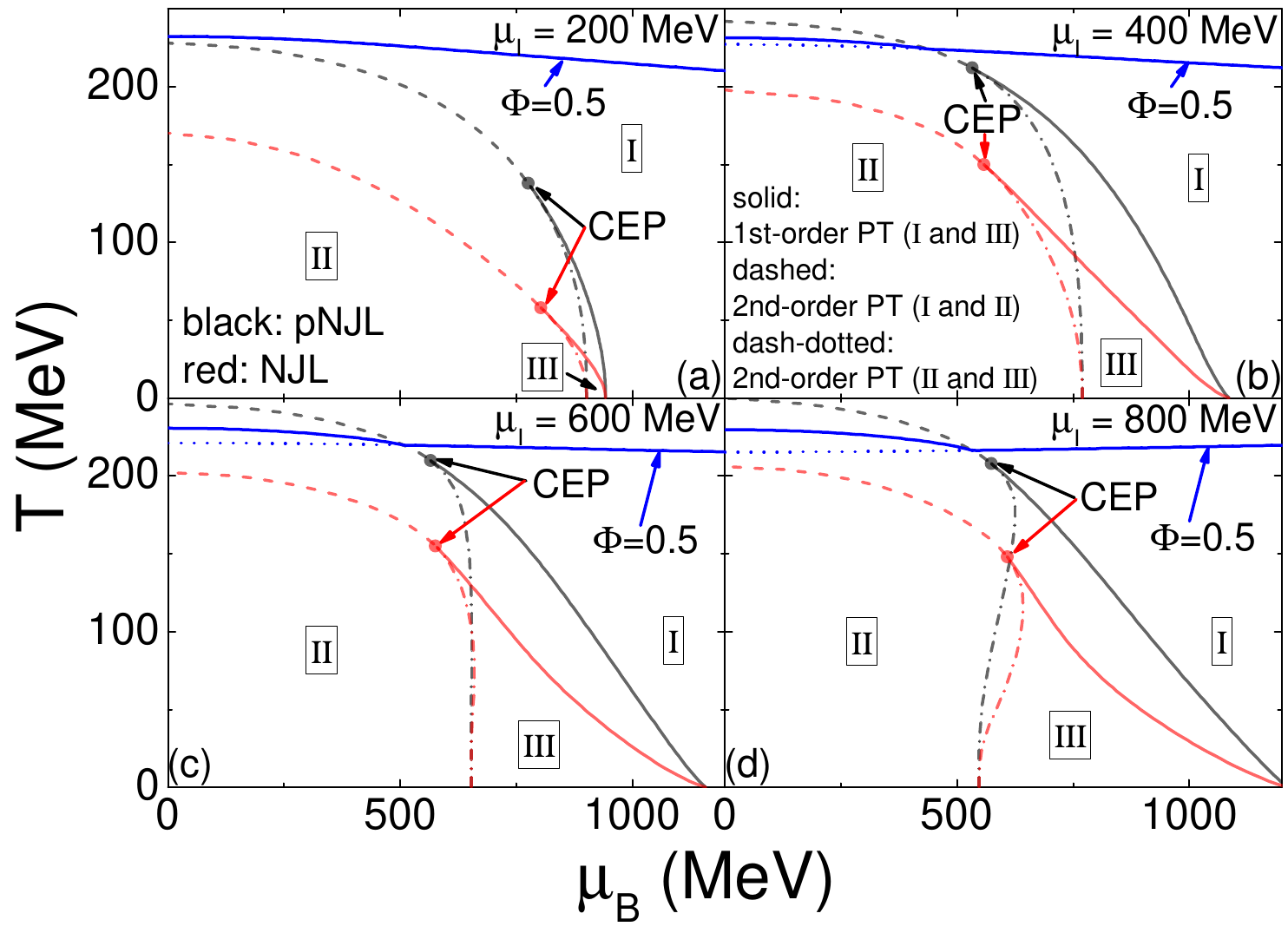}
	\caption{(Color online) Phase diagrams in the $T-\mu_\mathrm{B}$ plane at different isospin chemical potentials $\mu_\mathrm{I}=200$ (a), 400 (b), 600 (c), and 800 (d) MeV in the pNJL model compared with those in NJL model. Solid lines represent the first-order phase transition (PT) between Phase I and Phase III, dashed lines represent the second-order phase transition between Phase I and Phase II, and dash-dotted lines represent the second-order phase transition between Phase II and Phase III. Blue solid (dotted) lines represent the deconfinement phase transition with (without) the pion condensate. }\label{fig:PD-T-muB}
\end{figure}

\begin{figure}[ht]
	\includegraphics[scale=0.3]{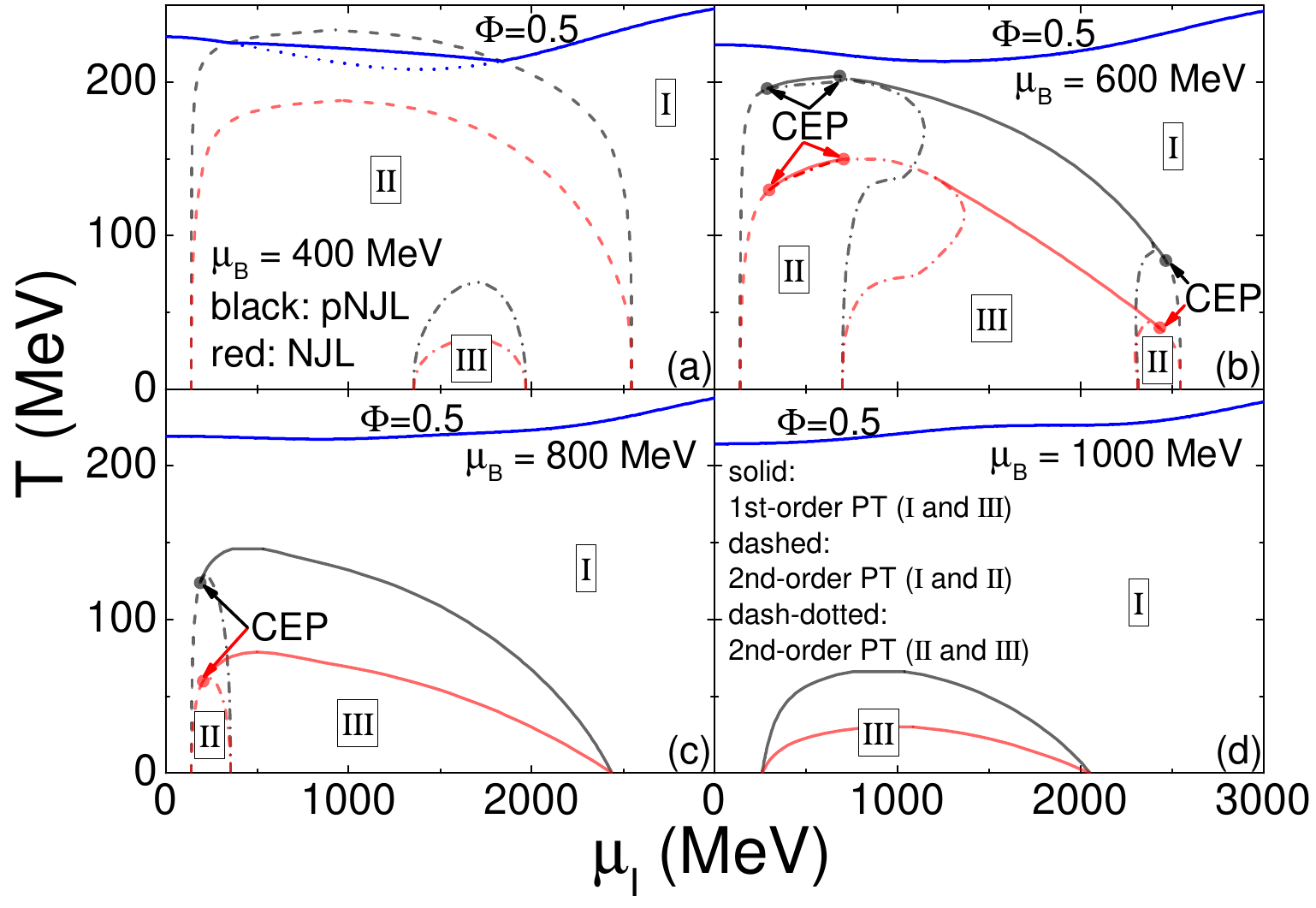}
	\caption{(Color online) Similar to Fig.~\ref{fig:PD-T-muB} but in the $T-\mu_\mathrm{I}$ plane at different baryon chemical potentials $\mu_\mathrm{B}=400$ (a), 600 (b), 800 (c), and 1000 (d) MeV.}\label{fig:PD-T-muI}
\end{figure}

\begin{figure}[ht]
	\includegraphics[scale=0.3]{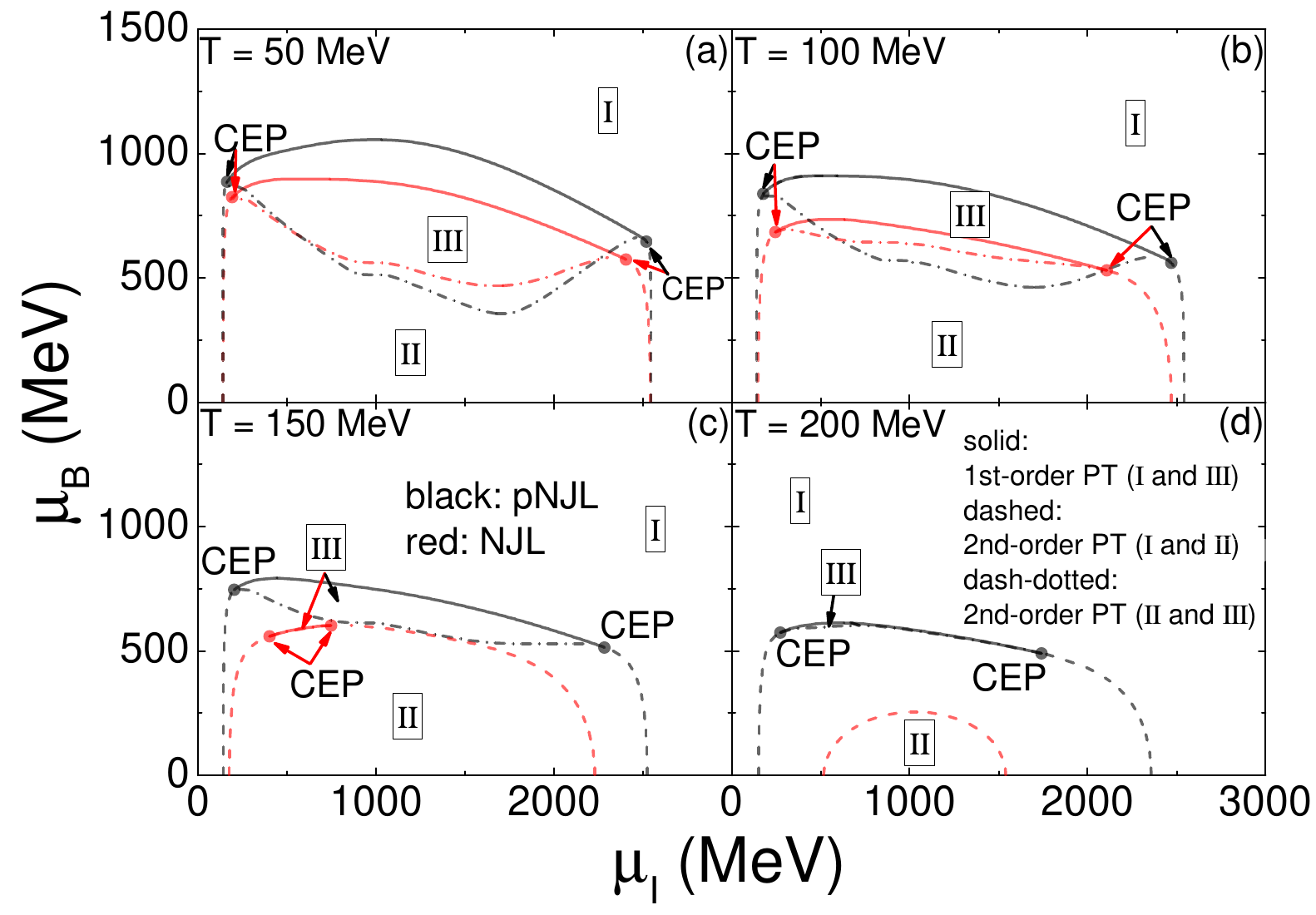}
	\caption{(Color online) Similar to Fig.~\ref{fig:PD-T-muB} but in the $\mu_\mathrm{B}-\mu_\mathrm{I}$ plane at different temperatures $T=50$ (a), 100 (b), 150 (c), and 200 (d) MeV.}\label{fig:PD-muB-muI}
\end{figure}

We now compare the three-dimensional ($T$, $\mu_\mathrm{B}$, $\mu_\mathrm{I}$) QCD phase diagram based on the pNJL model with that from the NJL model in Figs.~\ref{fig:PD-T-muB}, \ref{fig:PD-T-muI}, and \ref{fig:PD-muB-muI}, where areas of the normal baryon-rich and isospin-asymmetric quark matter with $\pi=0$ (Phases I), the pion superfluid phase with $\pi\ne0$ (Phase II), and the phase with both nonzero solutions of $\pi$ and $\pi_2$ (Phase III) as well as the corresponding phase boundaries will be presented. The transitions between Phase I and Phase III are always a first-order one indicated by the solid line, while the transitions between Phase I and Phase II as well as those between Phase II and Phase III are always a second-order one indicated by the dashed lines and dash-dotted lines, respectively. The CEPs are generally the crossing point for the three phases.

Figure~\ref{fig:PD-T-muB} displays the phase diagrams in the $T-\mu_\mathrm{B}$ plane at different isospin chemical potentials. For both NJL and pNJL models, Phase I generally exists at large $T$ or large $\mu_\mathrm{B}$, while Phase II generally exists at small $T$ and $\mu_\mathrm{B}$. The CEP, which connects the boundaries of the first-order phase transition and the second-order phase transitions, moves to a higher temperature with $\mu_\mathrm{I}$ changing from 200 MeV to 400 MeV, and the increasing trend saturates above $\mu_\mathrm{I}=400$ MeV. Compared to the NJL model, the pNJL model generally leads to larger areas of the pion superfluid phase and the Sarma phase, and a higher temperature of the CEP. Although the deconfinement phase transition in the pNJL model is always a smooth crossover, we plot an approximate deconfinement phase boundary for $\Phi=0.5$ with blue solid lines, and at smaller $\mu_\mathrm{B}$ it moves slightly to lower temperatures if there is no pion condensate as shown by blue dotted lines.

Figure~\ref{fig:PD-T-muI} displays the phase diagrams in the $T-\mu_\mathrm{I}$ plane at different baryon chemical potentials. For both NJL and pNJL models, the normal quark phase (Phase I) generally exists at very small or large isospin chemical potentials, or at high temperatures, while the area of the pion superfluid phase (Phase II) shrinks dramatically with the increasing baryon chemical potential. The phase transitions are always of second-order at small baryon chemical potentials, while the first-order phase transition becomes more and more dominate with the increasing baryon chemical potential. Phase III with $\pi_2 \ne 0$ doesn't exist at $\mu_\mathrm{B}=0$ (not shown here), but it gradually appears inside Phase II at small baryon chemical potentials, and its area becomes larger and dominates at large baryon chemical potentials. Similar to that in the $T-\mu_\mathrm{B}$ plane, the pNJL model leads to larger areas of the pion superfluid and Sarma phases and higher temperatures of the CEPs compared to the NJL model. The deconfinement phase transition happens at high temperatures, and can be affected by the pion condensate at smaller $\mu_\mathrm{B}$ and moderate $\mu_\mathrm{I}$.

Figure~\ref{fig:PD-muB-muI} displays the phase diagrams in the $\mu_\mathrm{B}-\mu_\mathrm{I}$ plane at different temperatures. For both NJL and pNJL models, the normal quark phase (Phase I) exists at larger $\mu_\mathrm{B}$ and/or very small or large $\mu_\mathrm{I}$, and the pion superfluid phase (Phase II) is observed at smaller $\mu_\mathrm{B}$ and moderate $\mu_\mathrm{I}$, already seen in Figs.~\ref{fig:PD-T-muB} and \ref{fig:PD-T-muI}. The area of Phase II shrinks with the increasing temperature. Also, the first-order phase transition and Phase III become less dominate at higher temperatures. Similarly, the pNJL model leads to larger areas of the pion superfluid and Sarma phases and CEPs at larger baryon chemical potentials compared to the NJL model.

\section{Summary and outlook}
\label{sec:summary}
To summarize, by introducing the gauge field and the Polyakov effective potential into the Lagrangian density of the extended three-flavor NJL model, we have studied the interplay among the chiral condensate, the pion condensate, and the Polyakov loop at finite temperatures, baryon chemical potentials, and isospin chemical potentials, and compared the three-dimensional QCD phase diagrams obtained from the NJL and pNJL models. While the two models give qualitatively similar QCD phase structures, we found that the pNJL model generally leads to larger areas of the pion superfluid and the Sarma phases and the CEP points at higher temperatures or larger baryon chemical potentials. The present study provides a more reliable prediction of the three-dimensional QCD phase diagram compared to our previous study without including the gluon dynamics.


As is well-known, NJL-type models are not normalizable and a momentum cutoff is generally needed to avoid divergence in the integral. This may affect the behavior of the model at larger chemical potentials, e.g., that of the Polyakov loop as found in the present study. The Pauli-Villars regularization scheme~\cite{Florkowski:1993bq,Mu:2008ic} could be a possible improvement for this drawback, and can be further investigated in future studies.

\begin{acknowledgments}
We acknowledge helpful discussions with Zhen-Yan Lu. Jun Xu is supported by the Strategic Priority Research Program of the Chinese Academy of Sciences under Grant No. XDB34030000 and the National Natural Science Foundation of China under Grant No. 12375125. Guang-Xiong Peng and Lu-Meng Liu are supported by the National Natural Science Foundation of China under Grant Nos. 11875052, 11575190, and 11135011.
\end{acknowledgments}

\bibliography{pion-pNJL}
\end{document}